\pdfoutput=1
\documentclass[11pt,a4paper]{article}
\usepackage[utf8]{inputenc}
\usepackage[T1]{fontenc}
\usepackage{lmodern}
\usepackage{microtype}
\usepackage[a4paper,margin=2.5cm]{geometry}
\usepackage[numbers,sort&compress]{natbib}
\usepackage{amsmath,amssymb,amsthm}
\usepackage{booktabs,array}
\usepackage{graphicx}
\usepackage{tikz-cd}
\usepackage[colorlinks=true,linkcolor=blue!50!black,citecolor=blue!50!black,urlcolor=blue!50!black]{hyperref}

\hypersetup{pdftitle={Generalization of the Theory of Relativity Based on Finsler Geometry, Incorporating a Fundamental Mass at High Energies},pdfauthor={Rustam Ibadov}}

\newtheorem{proposition}{Proposition}
\newtheorem{corollary}{Corollary}
\numberwithin{equation}{section}
\begin{document}

\title{\bfseries Generalization of the Theory of Relativity Based on Finsler Geometry, Incorporating a Fundamental Mass at High Energies}
\author{Rustam Ibadov\\[4pt]
{\small\itshape Department of Theoretical Physics and Quantum Electronics, Institute of Engineering Physics,}\\
{\small\itshape Samarkand State University, Samarkand, Uzbekistan}\\
{\small E-mail: \href{mailto:ibrustam@samdu.uz}{ibrustam@samdu.uz}}}
\date{}
\maketitle

\begin{abstract}
We generalize special and general relativity so that, besides the speed of light $c$, an invariant fundamental mass $M$ bounds the energy of every massive particle. The exact Legendre transform of a projectively deformed mass shell gives, for a particle of mass $m$, a closed-form Lorentzian Randers metric that depends on the ratio $m/M$. Its indicatrix is the Minkowski hyperboloid rescaled and shifted along a timelike direction, so the metric is of Zermelo navigation type, and its symmetry group, a shifted Lorentz group, realizes a relativity principle with two invariants, $c$ and $M$. Light propagates exactly as in Einstein's theory, while massive particles have energies below $M$, a deformed time dilation and a composition-dependent free fall. The Finsler--Einstein tensor equation, evaluated exactly along the comoving direction of a spatially flat universe, gives modified Friedmann equations and leaves de Sitter space exact. Particle energies, equivalence-principle tests and electron electric dipole moment limits constrain $M$. The construction makes Kadyshevsky's fundamental-mass hypothesis geometric and testable.

\end{abstract}

\section{Introduction}

Special relativity contains one invariant scale, the speed of light. Since Snyder's quantized spacetime~\cite{Snyder} there have been repeated attempts to add a second one, a fundamental length $\ell$ or, equivalently, a fundamental mass $M=\hbar/(\ell c)$. Gol'fand~\cite{Golfand} and later Kadyshevsky and collaborators~\cite{Kadyshevsky78,KMMV79} realized this idea by curving momentum space: the four-momentum is placed on a hyperboloid of radius $M$, de~Sitter in Gol'fand's construction and anti-de~Sitter in Kadyshevsky's, where the mass spectrum of elementary particles is bounded, $m\le M$. In the series ``Quantum field theory and a new universal high-energy scale'' this programme was carried through for scalar~\cite{KM85}, gauge vector~\cite{CDIKM85a,CDIKMprep} and Dirac~\cite{CDIKM85b} fields, with a local formulation in a five-dimensional configuration space. A new formulation of the theory and its unusual (exotic) fermion sector were developed in~\cite{IK89,DIKMC}, and its experimental consequences in~\cite{Ibadov04}. An algebraic, pseudo-Hermitian formulation of the maximal-mass Dirac theory was given in~\cite{RK16} and has since been quantized~\cite{Kravtsova21}. The programme was extended towards a maximal-mass generalization of the Standard Model~\cite{KMRS07} and to non-Hermitian $\gamma^5$ mass extensions, in which the bound $M=m_1^2/(2m_2)$ emerges algebraically and heavy exotic fermions are discussed as dark-matter candidates~\cite{RM17}. Markov's ``maximon'' hypothesis~\cite{Markov} reached a similar conclusion from gravitational arguments, identifying the limiting mass with the Planck mass.

A second line of work, doubly (deformed) special relativity (DSR)~\cite{AC02,MS02,MS04}, demands that the new scale be observer-independent, just as $c$ is. The Lorentz group is then realized nonlinearly on momentum space, and the dispersion relation of a free particle is modified at energies comparable to $M$. The Magueijo--Smolin (MS) realization~\cite{MS02} makes $M$ a maximal energy and is projective: the deformed boosts are fractional-linear transformations, in the spirit of Fock's analysis of the general form of relativistic transformations~\cite{Fock} and of the Beltrami coordinates Kadyshevsky used on curved momentum space. Curved momentum space is again an active topic, both as a route to well-defined quantum field theory~\cite{FV25,AKG21} and in the phenomenology of Planck-scale time dilation~\cite{AGPRR25,MLPAB24}; the observational status of deformed and broken relativistic symmetries is reviewed in~\cite{Addazi22}.

A modified dispersion relation is a statement about momentum space; its spacetime counterpart is a velocity-dependent, that is Finsler, geometry. Finsler spacetimes date back to Randers~\cite{Randers} and Bogoslovsky~\cite{Bogoslovsky}; Gibbons, Gomis and Pope showed that very special relativity~\cite{CG06} is Finsler geometry~\cite{GGP07}, and Girelli, Liberati and Sindoni showed that Planck-scale dispersion relations generically produce mass-dependent Finsler structures~\cite{GLS07}. Further developments include Lorentz-violating Finsler kinematics~\cite{Kostelecky11}, the geometry of physical dispersion relations~\cite{RRS11}, Hamilton geometry~\cite{BBGLP15}, Finsler realizations of DSR at first order~\cite{ABGLL14}, and Finsler gravity~\cite{Rutz,PW12,HPV19}. An alternative, algebraic route to Finsler extensions of relativity, based on poly-numbers and the Berwald--Moor metric, was outlined in~\cite{PPI08}. The modern foundations of Finsler spacetimes, namely cone structures and Lorentz--Finsler metrics~\cite{JS20}, observers~\cite{BJS20}, and action-based field theory~\cite{HPV22}, are reviewed in~\cite{LP18,Pf19}; Berwald spacetimes were related to very special relativity in~\cite{FPP18}, and Schwarzschild--Finsler--Randers spacetimes, of the same $\alpha+\beta$ type as ours, were studied in~\cite{TBKS20,KTBS21,KKST22,TKS24}. Recent work has studied $(\alpha,\beta)$-metrics and exact vacuum solutions in Finsler gravity~\cite{Heefer24}, classified Einstein--Kropina solutions~\cite{HVF26}, clarified when the Finsler vacuum equation reduces to Ricci-flatness~\cite{PV26}, and applied Finsler, Randers and Kropina geometries to cosmology~\cite{HPV20c,PVFP25,Bouali25}. Rodrigues and Lobo stated conditions under which a mass level set of a phase-space Hamiltonian, i.e.\ a modified dispersion relation, defines a Finsler function whose geodesics are the projected Hamiltonian trajectories, and noted that not every Hamiltonian admits such a transformation~\cite{RL24}; the projective mass shell of Sec.~\ref{sec:legendre} is a case in which the transformation exists and is given in closed form.

In this paper we combine these strands and obtain an exact, closed-form Finsler geometry in which $M$ is a fundamental mass. Our main results are:
\begin{enumerate}
\item The Legendre dual of the projective mass shell is a Lorentzian Randers metric $F_\mu$, $\mu=m/M$, given in closed form (Sec.~\ref{sec:legendre}).
\item $F_\mu$ has Lorentzian signature inside the timelike cone, its null cone is the light cone of $g$, and it interpolates between Minkowski ($\mu\to0$) and Kropina ($\mu\to1$) geometry (Sec.~\ref{sec:geometry}).
\item The indicatrix of $F_\mu$ is a displaced Minkowski hyperboloid, and the relativity group is the Lorentz group centred at the displacement $W=\mu u/(1+\mu)$. This demonstrates explicitly that a relativity principle with two invariants, $c$ and $M$, exists (Sec.~\ref{sec:demo}).
\item Massive particles cannot exceed the energy $M$, and all rest energies obey $\varepsilon_0<M$ (Sec.~\ref{sec:kinematics}).
\item In curved spacetime the deformation acts as an ``aether charge'' $e\simeq m^2/M$, giving a definite violation of the weak equivalence principle (Sec.~\ref{sec:curved}), which together with laboratory tests constrains $M$ (Sec.~\ref{sec:constraints}).
\item The Finsler--Einstein tensor equation $G_{ab}(x,y)=\kappa T_{ab}(x,y)$, built from the Akbar-Zadeh Ricci tensor of $F_\mu$, has the same vacuum solutions as Einstein's equation when the frame field is parallel. On a spatially flat FLRW background we evaluate it exactly along the comoving direction and obtain modified Friedmann equations (Sec.~\ref{sec:fieldeq}).
\end{enumerate}
We use units $\hbar=c=1$ and signature $(+,-,-,-)$. Indices $a,b,\dots$ run over $0$--$3$; a dot denotes the derivative with respect to an arbitrary curve parameter $\lambda$.

\section{Fundamental mass as an invariant scale}\label{sec:mass}

\subsection{Kadyshevsky's geometric approach}\label{sec:kadyshevsky}

In the quantum field theory with a fundamental mass, $M$ is the curvature radius of four-dimensional momentum space~\cite{Kadyshevsky78,KM85}. The four-momentum lies on the hyperboloid
\begin{equation}
p_0^2-\mathbf p^2+p_5^2=M^2,\qquad \varepsilon=\frac{p_5}{|p_5|}=\pm1,
\label{eq:ads}
\end{equation}
where the sign $\varepsilon$ is a new discrete quantum number. On the mass shell $p_5=\varepsilon M\cos\chi$ with $\cos\chi=\sqrt{1-m^2/M^2}$, so the mass spectrum is bounded, $m\le M$. Objects heavier than $M$ cannot be elementary, because no local field corresponds to them; $M$ plays the role of Markov's maximon mass~\cite{Markov,RK16}.

\paragraph{Five-dimensional locality} Treating $p_5$ as the momentum conjugate to an extra coordinate $x^5$, the theory admits a local formulation in five-dimensional configuration space~\cite{KM85,CDIKM85a,CDIKM85b,Ibadov04}. The free field obeys
\begin{equation}
\left(\frac{\partial^2}{\partial x^\mu\partial x_\mu}-\frac{\partial^2}{\partial x_5^2}-M^2\right)\Phi(x,x^5)=0,
\label{eq:5d}
\end{equation}
which corresponds to the de~Sitter hyperboloid $p^2-p_5^2=-M^2$ used in~\cite{Ibadov04}; the anti-de~Sitter case~\eqref{eq:ads} follows by reversing the signs of both the $\partial^2/\partial x_5^2$ and the $M^2$ terms. Its Cauchy data at $x^5=0$, $\Phi(x,0)$ and $\partial\Phi(x,0)/\partial x^5$, double the number of field components in ordinary spacetime. The action $S=\int d^4x\,L[\Phi(x,0),\partial\Phi(x,0)/\partial x^5]$ is required to satisfy $\partial S/\partial x^5=0$, and the flat limit $M\to\infty$ returns standard quantum field theory.

\paragraph{Gauge fields} For the electromagnetic and Yang--Mills fields~\cite{CDIKM85a,CDIKMprep}, the gauge symmetry of this formulation resembles a local five-dimensional gauge symmetry broken by a universal mechanism. Gauge invariance then generates interaction vertices proportional to $\gamma^5$ and $\gamma^5\sigma_{\mu\nu}$, whose strength is set by $M$. These vertices violate P and CP, do not conserve helicity at ultrahigh energies, and give leptons an electric dipole moment $d\simeq e\ell/2$ with $\ell=1/M$~\cite{Ibadov04}. A gauge-invariant (Stueckelberg) massive vector field reduced from a higher-dimensional space produces a closely related nonlocal square-root kernel in recent work on planar QED~\cite{BAN26}; we return to this analogy in Sec.~\ref{sec:discussion}.

\paragraph{Fermions} Non-Euclidean momentum space admits several equally valid Dirac operators~\cite{CDIKM85b}. The normal and the exotic one are
\begin{equation}
\mathcal D=p_n\gamma^n-(p_5-M)\gamma^5+2M\sin\tfrac{\chi}{2},\qquad
\mathcal D_{\rm exotic}=p_n\gamma^n+(p_5+M)\gamma^5-2M\cos\tfrac{\chi}{2}.
\end{equation}
Exotic fields grow as $\sqrt M$ in the flat limit and have no counterpart in ordinary field theory. They form the ``unusual fermion matter'' of~\cite{DIKMC}; their non-Hermitian mass structure was later developed in a pseudo-Hermitian algebraic form~\cite{RK16}.

\subsection{Projective realization}\label{sec:projective}

Kadyshevsky's free mass shell is still $p^2=m^2$, so its Legendre dual is the pseudo-Riemannian norm, and all new physics lives in the interactions and in the doubled field content. To obtain a nontrivial Finsler geometry of spacetime we keep the central idea, an upper bound set by $M$, but replace the linear Lorentz invariance of the hyperboloid by a nonlinear realization.

Let $u$ be the unit timelike covector of an inertial observer, $u\cdot u=1$, and let $p_a$ be the momentum that observer measures, with energy $E=u^ap_a$. We introduce the projective map
\begin{equation}
U:\; p_a\longmapsto \pi_a=\frac{p_a}{1-u^bp_b/M},\qquad
p_a=\frac{\pi_a}{1+u^b\pi_b/M}.
\label{eq:U}
\end{equation}
The auxiliary momenta $\pi$ transform linearly under the Lorentz group $\Lambda$. The physical momenta transform under the conjugate group $\Lambda_M=U^{-1}\Lambda U$, which is nonlinear and fractional-linear in $p$. Under $\Lambda_M$ the quantities $c$ and $M$ are both invariant: the map fixes the null cone $E=|\mathbf p|$ and the hyperplane $E=M$.

The mass shell is the Casimir of $\Lambda_M$,
\begin{equation}
\mathcal C_M(p)\equiv\frac{\eta^{ab}p_ap_b}{(1-u^cp_c/M)^2}=m^2,
\label{eq:casimir}
\end{equation}
which is the Magueijo--Smolin relation~\cite{MS02}. Its physical branch is the image of the future sheet $\pi_0>0$ of the ordinary hyperboloid; on it
\begin{equation}
E=\frac{\pi_0}{1+\pi_0/M}<M.
\end{equation}
The energy of every massive particle is therefore bounded by $M$. At rest ($\pi_0=m$) the physical rest energy is
\begin{equation}
\varepsilon_0=\frac{mM}{M+m},\qquad \frac{1}{\varepsilon_0}=\frac1m+\frac1M,
\label{eq:eps0}
\end{equation}
so that $\varepsilon_0<M$ for any Casimir mass $m$, with $\varepsilon_0\to M$ only as $m\to\infty$. This is the kinematic content of Kadyshevsky's fundamental-mass hypothesis~\cite{Kadyshevsky78,KMMV79}: no elementary object can carry a rest energy above $M$. In Kadyshevsky's anti-de~Sitter momentum space the bound is $m\le M$ and is reached at finite $m$; here it is approached asymptotically, and $M$ plays the role of a ``reduced mass'' partner of every particle. Massless particles ($m=0$) satisfy $E=|\mathbf p|$ exactly.

The parameter that controls all deviations from special relativity is
\begin{equation}
\mu\equiv\frac mM,\qquad \varepsilon_0=\frac{m}{1+\mu}.
\end{equation}

\section{From the mass shell to a Finsler function}\label{sec:legendre}

The reparametrization-invariant action of a free particle in phase space is
\begin{equation}
S=-\int d\lambda\,\Big[p_a\dot x^a-\tfrac{\nu}{2}\big(\mathcal C_M(p)-m^2\big)\Big].
\end{equation}
Eliminating $p$ and the multiplier $\nu$ gives $S=-\int L(\dot x)\,d\lambda$, where $L(\dot x)$ is the stationary value of $p_a\dot x^a$ on the mass shell~\cite{GLS07,RRS11,RL24}. Writing $p$ through $\pi$ and using $\pi^2=m^2$, we must find the stationary value of
\begin{equation}
f(\pi)=\frac{\pi\cdot\dot x}{1+u\cdot\pi/M}\qquad\text{on}\quad \pi\cdot\pi=m^2,\ \pi_0>0.
\end{equation}

\paragraph{Stationarity} Varying $f$ with a Lagrange multiplier shows that $\pi$ is parallel to $w=\dot x-a\,u$ with $a=f/M$, hence $\pi=m\,w/\sqrt{w\cdot w}$. Inserting this into $f$ and writing $R=\sqrt{w\cdot w}$, the condition $f=Ma$ reduces after cancellation to the single relation $a=\mu R$. Squaring gives
\begin{equation}
(1-\mu^2)\,a^2+2\mu^2Ta-\mu^2(T^2-X^2)=0,\qquad T=u\cdot\dot x,\quad X^2=T^2-\dot x\cdot\dot x,
\end{equation}
whose root with $a\ge0$ is the physical one. It exists precisely for timelike or null $\dot x$, i.e.\ for velocities $v=X/T\le1$.

\paragraph{Result} The Lagrangian is $L=\varepsilon_0F_\mu(\dot x)$, where $F_\mu$ is normalized so that $F_\mu(u)=1$:
\begin{equation}
\boxed{\;F_\mu(\dot x)=\frac{\sqrt{(1-\mu^2)\,\eta_{ab}\dot x^a\dot x^b+\mu^2(u_a\dot x^a)^2}\;-\;\mu\,u_a\dot x^a}{1-\mu}\;}
\label{eq:F}
\end{equation}
The free-particle action is $S=-\varepsilon_0\int F_\mu(\dot x)\,d\lambda$. The function $F_\mu$ is positively homogeneous of degree one, so $S$ is reparametrization invariant and $F_\mu$ is a Finsler function on the future timelike cone. Since it depends on $m$ through $\mu$, particles of different mass probe different Finsler geometries, as anticipated in~\cite{GLS07}.

\paragraph{Checks} (i) For $\mu\to0$, $F_\mu\to\sqrt{\eta_{ab}\dot x^a\dot x^b}$: special relativity. (ii) For a particle at rest $L=\varepsilon_0T$, so the canonical energy at rest is $\varepsilon_0$, consistent with Eq.~\eqref{eq:eps0}. (iii) The canonical momenta $p_a=\varepsilon_0\,\partial F_\mu/\partial\dot x^a$ satisfy the Casimir relation~\eqref{eq:casimir} identically. We verified (iii) analytically and numerically to $10^{-10}$ for $0<\mu<1$ and $0\le v<1$.

\paragraph{Symmetry} The linear isometries of $F_\mu$ in a tangent space are the rotations that fix $u$. The full six-parameter group $\Lambda_M$ acts on velocities through the Legendre map, which is a diffeomorphism because the fundamental tensor is nondegenerate (Sec.~\ref{sec:geometry}). In Sec.~\ref{sec:demo} we show that this action is affine on the indicatrix and construct it explicitly. Amelino-Camelia \emph{et al.}~\cite{AGPRR25} stress that a Finsler time-dilation law is DSR-relativistic only if the line element is invariant under the deformed boosts. Morais \emph{et al.}~\cite{MLPAB24} showed that the relativity principle and the clock postulate are compatible in DSR once the deformed Lorentz transformations are written in position space. Proposition~\ref{prop:group} shows that $F_\mu$ is invariant under the deformed boosts induced on the tangent space, for every species; the residual issue, the species dependence of the transformation, is quantified in Sec.~\ref{sec:species}.

\section{Geometry of \texorpdfstring{$F_\mu$}{F\_mu}}\label{sec:geometry}

\paragraph{$(\alpha,\beta)$ and Randers form} With $\alpha=\sqrt{g_{ab}\dot x^a\dot x^b}$, $\beta=u_a\dot x^a$ and $s=\beta/\alpha\ge1$ on the future timelike cone, $F_\mu$ is an $(\alpha,\beta)$-metric~\cite{BCS00}, $F_\mu=\alpha\,\phi(s)$, with
\begin{equation}
\phi(s)=\frac{Q(s)-\mu s}{1-\mu},\qquad Q(s)=\sqrt{1-\mu^2+\mu^2s^2}.
\end{equation}
Equivalently, $F_\mu$ is a Lorentzian Randers metric~\cite{Randers}, $F_\mu=\tilde\alpha+\tilde\beta$, with
\begin{equation}
\tilde a_{ab}=\frac{(1-\mu^2)\,g_{ab}+\mu^2u_au_b}{(1-\mu)^2},\qquad
\tilde\beta=-\frac{\mu}{1-\mu}\,u_a\,dx^a.
\label{eq:randers}
\end{equation}
In the rest frame of $u$, $\tilde a=(1-\mu)^{-2}dt^2-(1+\mu)(1-\mu)^{-1}d\mathbf x^2$, whose light cone has the speed $(1-\mu^2)^{-1/2}>1$. For $\mu>1$ the tensor $\tilde a$ is positive definite and, since $1-\mu<0$, the same expression~\eqref{eq:F} takes the form $F_\mu=\tilde\beta-\tilde\alpha$ with $\tilde\alpha=\sqrt{\tilde a_{ab}\dot x^a\dot x^b}$ (last row of Table~\ref{tab:limits}). The one-form $\tilde\beta$ closes that wider cone back onto the light cone of $g$: $F_\mu$ vanishes exactly when $g_{ab}\dot x^a\dot x^b=0$. The causal structure of the theory is therefore that of the metric $g$, which is the geometric reason why photons are unaffected (Sec.~\ref{sec:constraints}).

\paragraph{Fundamental tensor} The Finsler metric $g^F_{ab}=\tfrac12\,\partial^2F_\mu^2/\partial\dot x^a\partial\dot x^b$ follows from the general $(\alpha,\beta)$ formulas~\cite{BCS00}. We need
\begin{equation}
\phi-s\phi'=\frac{1+\mu}{Q},\qquad \phi''=\frac{\mu^2(1+\mu)}{Q^3},\qquad
(\phi-s\phi')+(b^2-s^2)\,\phi''=\frac{1+\mu}{Q^3},
\end{equation}
where $b^2=g^{ab}u_au_b=1$. The determinant in four dimensions is
\begin{equation}
\det g^F_{ab}=\left[\frac{(1+\mu)^3\,\phi^5}{Q^5}\right]\det g_{ab}.
\label{eq:det}
\end{equation}

\begin{proposition}\label{prop:signature}
For every $\mu\ge0$ (for $\mu=1$ through the regular form~\eqref{eq:Fregular} below), $g^F_{ab}$ is nondegenerate with Lorentzian signature $(+,-,-,-)$ on the open future timelike cone of $g$, and degenerates on its boundary.
\end{proposition}
\begin{proof}
The bracket in~\eqref{eq:det} is strictly positive for $\phi>0$, i.e.\ inside the cone. At $\mu=0$, $g^F=g$; since the determinant does not vanish as $\mu$ varies, no eigenvalue changes sign. On the light cone $\phi\to0$.
\end{proof}

We confirmed Eq.~\eqref{eq:det} by direct numerical differentiation (relative agreement $3\times10^{-9}$). The degeneracy on the light cone is shared with the Bogoslovsky and Kropina metrics~\cite{Bogoslovsky,GGP07}. $F_\mu$ therefore defines a Finsler spacetime in the sense of a smooth cone structure with a Lorentzian Finsler norm on the open timelike cone~\cite{Beem,JS20}.

\paragraph{Limits} The one-parameter family connects two classical geometries (Table~\ref{tab:limits}). The equivalent form
\begin{equation}
F_\mu=\frac{(1+\mu)\,\alpha^2}{\sqrt{(1-\mu^2)\alpha^2+\mu^2\beta^2}+\mu\beta}
\label{eq:Fregular}
\end{equation}
shows that the limit $\mu\to1$ is regular.

\begin{table}[h]
\centering
\caption{Limits of the Finsler function $F_\mu$.}\label{tab:limits}
\small
\begin{tabular}{@{}lll@{}}
\toprule
$\mu=m/M$ & $F_\mu$ & Geometry\\
\midrule
$0$ & $\alpha$ & pseudo-Riemannian (Einstein)\\
$0<\mu<1$ & $\big(\sqrt{(1-\mu^2)\alpha^2+\mu^2\beta^2}-\mu\beta\big)/(1-\mu)$ & Lorentzian Randers, $\tilde\alpha+\tilde\beta$\\
$1$ & $\alpha^2/\beta$ & Kropina\\
$>1$ & $\big(\mu\beta-\sqrt{\mu^2\beta^2-(\mu^2-1)\alpha^2}\big)/(\mu-1)$ & Randers type, $\tilde\beta-\tilde\alpha$\\
\bottomrule
\end{tabular}
\end{table}

\section{Why the generalization exists: a geometric demonstration}\label{sec:demo}

The fundamental mass does not destroy the relativistic structure of the space of velocities; it only displaces it. In this section we show that the unit sphere of $F_\mu$ is the Minkowski hyperboloid moved by a fixed vector, and that the relativity group is the Lorentz group rotating about the displaced centre. This makes the existence of a relativity theory with two invariant scales, $c$ and $M$, explicit. All statements below were also verified numerically, to a relative accuracy of $10^{-12}$, for random $0<\mu<1$, velocities and boosts.

\subsection{The indicatrix is a displaced hyperboloid}

Let $\mathcal H=\{w:\ g(w,w)=1,\ u\cdot w>0\}$ be the unit hyperboloid of special relativity, and write $u\cdot w=\cosh\xi$, where $\xi\ge0$ is the rapidity of $w$ relative to $u$.

\begin{proposition}[Displaced indicatrix]\label{prop:indicatrix}
For every $w\in\mathcal H$, $F_\mu(w+\mu u)=1+\mu$. Hence the indicatrix $I_\mu=\{X:\ F_\mu(X)=1\}$ is
\begin{equation}
I_\mu=\frac{\mathcal H+\mu u}{1+\mu}=W+\frac{\mathcal H}{1+\mu},\qquad W\equiv\frac{\mu}{1+\mu}\,u .
\label{eq:indicatrix}
\end{equation}
\end{proposition}
\begin{proof}
With $\gamma=u\cdot w=\cosh\xi$ one has $g(w+\mu u,w+\mu u)=1+2\mu\gamma+\mu^2$ and $u\cdot(w+\mu u)=\gamma+\mu$. The radicand in Eq.~\eqref{eq:F} becomes
$(1-\mu^2)(1+2\mu\gamma+\mu^2)+\mu^2(\gamma+\mu)^2=(1+\mu\gamma)^2$, so
$F_\mu(w+\mu u)=[(1+\mu\gamma)-\mu(\gamma+\mu)]/(1-\mu)=1+\mu$. The second statement follows from homogeneity.
\end{proof}

\begin{proposition}[Zermelo navigation]\label{prop:zermelo}
For future-directed $X$,
\begin{equation}
\begin{gathered}
F_\mu(X)=1\quad\Longleftrightarrow\quad h(X-W,\,X-W)=1,\\
h=(1+\mu)^2g,\qquad h(W,W)=\mu^2 .
\end{gathered}
\label{eq:zermelo}
\end{equation}
\end{proposition}
This is the Lorentzian form of Zermelo's navigation problem~\cite{BRS04}: the Randers metric~\eqref{eq:randers} is the navigation metric of the ``sea'' $h$ in the presence of the ``wind'' $W$~\cite{GHWW09,CJS24}. The fundamental mass enters the geometry only as a wind of strength $\mu$ along the observer's time direction. The critical case $h(W,W)=1$, in which $W$ lies on the indicatrix of $h$, is $\mu=1$; it gives the Kropina metric of Table~\ref{tab:limits}, in agreement with the role of Kropina metrics as critical-wind navigation metrics~\cite{CJS24}.

\begin{figure}[t]
\centering
\includegraphics[width=0.92\textwidth]{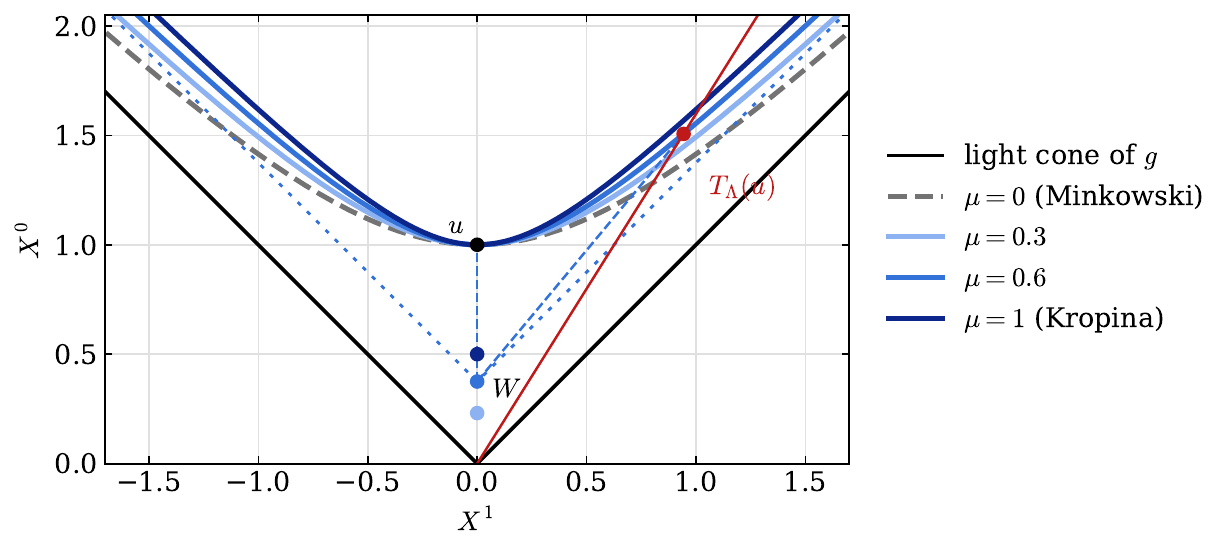}
\caption{Indicatrices $I_\mu=\{F_\mu=1\}$ in the plane spanned by $u$ and one spatial direction. Each is the Minkowski hyperbola (dashed) scaled by $1/(1+\mu)$ and displaced by the wind $W=\mu u/(1+\mu)$ (dots). All pass through $u$, and all have the light cone as the cone of their rays; the dotted lines are the asymptotes of $I_{0.6}$, the light cone moved to $W$. A boost with rapidity $\theta=1.2$ rotates $I_{0.6}$ about $W$ (dashed radii) and takes $u$ to $T_\Lambda(u)$. The physical velocity is the slope of the ray from the origin (red line): $v=0.63$, against $\tanh1.2=0.83$ in special relativity.}
\label{fig:indicatrix}
\end{figure}

The asymptotic cone of $I_\mu$ is the light cone of $g$ translated to $W$ (Fig.~\ref{fig:indicatrix}). Rays from the origin through the points of $I_\mu$ therefore fill exactly the open future timelike cone, and they become null only at infinity on $I_\mu$. A displacement of the indicatrix along $u$ cannot tilt its asymptotes, and this is the geometric reason why the speed of light stays invariant. Every indicatrix passes through $u$ itself, since $F_\mu(u)=1$: the state of rest is the same for all $\mu$.

\subsection{The Legendre map is affine}

Differentiating the Casimir~\eqref{eq:casimir},
\begin{equation}
\frac{\partial\mathcal C_M}{\partial p_a}=\frac{2}{(1-E/M)^2}\left[p^a+\frac{m^2}{M}\Big(1-\frac EM\Big)u^a\right]\ \propto\ \pi^a+\mu\,m\,u^a ,
\end{equation}
where we used $p=\pi/(1+u\cdot\pi/M)$ from Eq.~\eqref{eq:U}. The velocity of a particle with auxiliary momentum $\pi=m\,w$ is therefore proportional to $w+\mu u$, and the Legendre map sends the point $w=\pi/m$ of the unit hyperboloid to the point $A_\mu(w)=(w+\mu u)/(1+\mu)$ of $I_\mu$. The whole construction is summarized by the diagram
\begin{equation}
\begin{tikzcd}[column sep=5em, row sep=2.5em]
\Sigma_m \arrow[loop above, "\Lambda_M"] & \mathcal H \arrow[l, "U^{-1}"', "\text{projective}"] \arrow[r, "A_\mu"', "\text{affine}"] \arrow[loop above, "\Lambda"] & I_\mu \arrow[loop above, "G_\mu"]
\end{tikzcd}
\qquad
\mathcal L=A_\mu\circ U/m ,
\label{eq:diagram}
\end{equation}
in which $\Sigma_m$ is the physical mass shell (momentum space, $E<M$), $\mathcal H$ is the hyperboloid of special relativity, $I_\mu$ is the indicatrix (tangent space), and $\mathcal L:\Sigma_m\to I_\mu$ is the Legendre map; the arrow $U^{-1}$ is understood as $w\mapsto U^{-1}(mw)$, so that $\mathcal L(p)=A_\mu\big(U(p)/m\big)$. Special relativity sits in the middle; the fundamental mass appears as a projective deformation on the momentum side and as an affine deformation on the velocity side.

\subsection{The relativity group}

\begin{proposition}[Relativity group]\label{prop:group}
The affine transformations of the tangent space that map $I_\mu$ onto itself are
\begin{equation}
T_\Lambda:\ X\longmapsto W+\Lambda(X-W),\qquad \Lambda\in O^+(1,3),
\label{eq:TLambda}
\end{equation}
and they form a group isomorphic to the orthochronous Lorentz group $O^+(1,3)$; its identity component, $G_\mu\cong SO^+(1,3)$, contains the deformed boosts and rotations. Their homogeneous extensions
\begin{equation}
\hat T_\Lambda(X)=F_\mu(X)\,T_\Lambda\big(X/F_\mu(X)\big)
\end{equation}
\begin{enumerate}
\item[(i)] preserve $F_\mu$: $F_\mu(\hat T_\Lambda X)=F_\mu(X)$;
\item[(ii)] map the open future timelike cone onto itself;
\item[(iii)] extend continuously to the light cone, on which $\hat T_\Lambda=\Lambda$;
\item[(iv)] are the Legendre images of the deformed boosts: $\mathcal L\circ\Lambda_M=T_\Lambda\circ\mathcal L$.
\end{enumerate}
\end{proposition}
\begin{proof}
An affine map that preserves a nondegenerate quadric fixes its centre, here $W$, and its linear part preserves $h$, hence $g$. It must be orthochronous to preserve the future sheet; spatial reflections are allowed, which gives $O^+(1,3)$. (i) holds by construction, and (ii) because $\hat T_\Lambda$ maps rays bijectively onto rays. For (iii), as $X$ approaches the light cone, $F_\mu(X)\to0$ and $Y=X/F_\mu(X)$ runs to infinity on $I_\mu$, where $T_\Lambda(Y)=\Lambda Y+O(1)$; multiplying by $F_\mu(X)$ gives $\hat T_\Lambda(X)\to\Lambda X$. For (iv), $U$ intertwines $\Lambda_M$ with $\Lambda$ by definition, and $A_\mu(\Lambda w)=(\Lambda w+\mu u)/(1+\mu)=W+\Lambda\big(A_\mu(w)-W\big)=T_\Lambda\big(A_\mu(w)\big)$.
\end{proof}

In special relativity the boosts rotate the indicatrix about zero velocity. With a fundamental mass they rotate the displaced indicatrix about $W$. The group is unchanged, $G_\mu\cong SO^+(1,3)$ (or $O^+(1,3)$ with reflections), so the relativity principle survives; what changes is the point about which velocities are ``rotated''. Item~(iii) states that light is transformed exactly as in Einstein's theory. The invariance of $M$ follows because $\Lambda_M$ maps the mass shell $\Sigma_m$, on which $E<M$, onto itself. Both invariants are therefore built into the geometry: $c$ as the common asymptotic cone of all indicatrices, and $M$ as the bound of all mass shells.

\subsection{The deformation in one table}

Because $\Lambda$ acts on $\mathcal H$ by shifts of rapidity, $\xi$ is additive under collinear transformations of $G_\mu$: $\xi\mapsto\xi-\theta$. Every kinematic quantity follows from Proposition~\ref{prop:indicatrix} and Eq.~\eqref{eq:U}, and Table~\ref{tab:rapidity} compares them with special relativity. The expressions reproduce Eq.~\eqref{eq:Ep} and the time-dilation law of Sec.~\ref{sec:kinematics}.

\begin{table}[h]
\centering
\caption{Kinematics as a function of the additive rapidity $\xi$.}\label{tab:rapidity}
\small
\setlength{\tabcolsep}{4pt}
\begin{tabular}{@{}llll@{}}
\toprule
Quantity & Special relativity ($\mu=0$) & With fundamental mass & $\xi\to\infty$ \\
\midrule
Energy $E$ & $m\cosh\xi$ & $\dfrac{m\cosh\xi}{1+\mu\cosh\xi}$ & $M$\\[8pt]
Momentum $|\mathbf p|$ & $m\sinh\xi$ & $\dfrac{m\sinh\xi}{1+\mu\cosh\xi}$ & $M$\\[8pt]
Velocity $v$ & $\tanh\xi$ & $V_\mu(\xi)=\dfrac{\sinh\xi}{\cosh\xi+\mu}$ & $c=1$\\[8pt]
Clock rate $d\tau/dt$ & $1/\cosh\xi$ & $\dfrac{1+\mu}{\cosh\xi+\mu}$ & $0$\\[8pt]
Rest energy ($\xi=0$) & $m$ & $m/(1+\mu)$ & ---\\
\bottomrule
\end{tabular}
\end{table}

\begin{figure}[t]
\centering
\begin{minipage}{0.49\textwidth}\centering
\includegraphics[width=\textwidth]{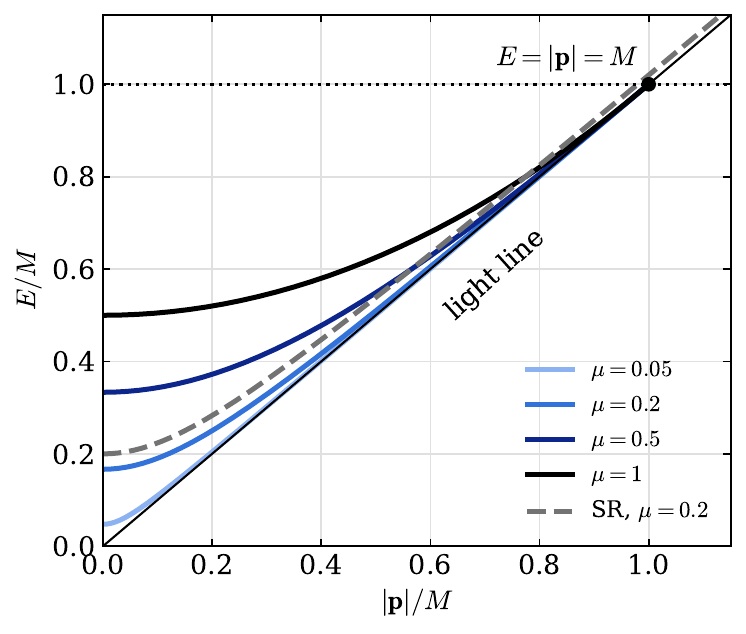}\\[2pt]
\small (a) Mass shells
\end{minipage}\hfill
\begin{minipage}{0.49\textwidth}\centering
\includegraphics[width=\textwidth]{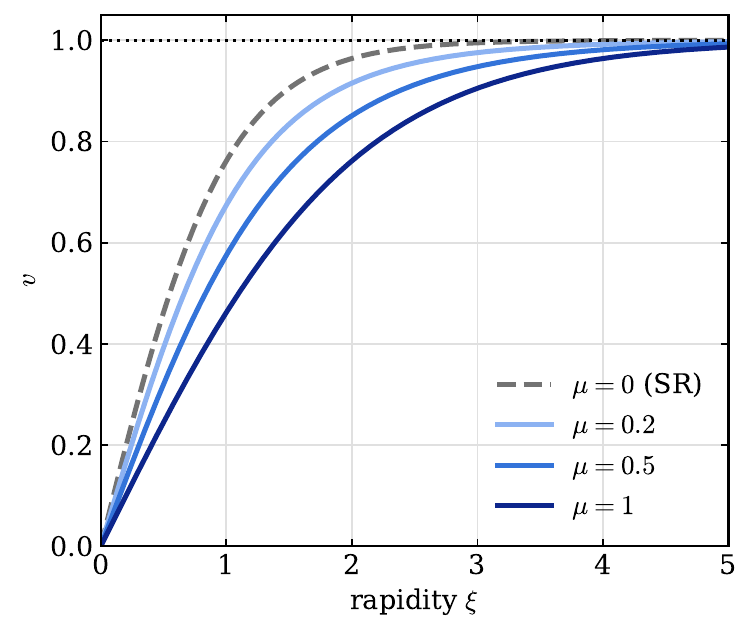}\\[2pt]
\small (b) Velocity versus rapidity
\end{minipage}
\caption{(a) Physical mass shells $\Sigma_m$ for several $\mu=m/M$ (solid). Every shell lies below $E=M$ and ends at the corner point $E=|\mathbf p|=M$ on the light line; the special-relativistic shell for $\mu=0.2$ (dashed) crosses $E=M$. (b) The measured velocity $V_\mu(\xi)$ as a function of the additive rapidity. The speed of light is reached only as $\xi\to\infty$, for every $\mu$.}
\label{fig:shells}
\end{figure}

Composition of collinear motions is addition of $\xi$. A particle that has velocity $v$ in one frame has, in a frame boosted by $\theta$,
\begin{equation}
v'=V_\mu\big(V_\mu^{-1}(v)-\theta\big),\qquad
\tanh\frac{V_\mu^{-1}(v)}{2}=\frac{1-\sqrt{1-(1-\mu^2)v^2}}{(1-\mu)\,v}.
\label{eq:composition}
\end{equation}
For $\mu=0$ this is Einstein's addition law. As $\xi\to\infty$, $v\to c$ and $E,|\mathbf p|\to M$ simultaneously: every mass shell ends at the corner point $E=|\mathbf p|=M$ on the light line (Fig.~\ref{fig:shells}).

\subsection{Several species: the size of the residual}\label{sec:species}

Because $W$ depends on $\mu$, particles of different mass are transformed by different groups $G_\mu$. Two particles at rest relative to each other in one frame ($\xi=0$ for both) acquire, in a frame boosted by $\theta$, the velocities $V_{\mu_1}(\theta)$ and $V_{\mu_2}(\theta)$, and
\begin{equation}
\big|V_{\mu_1}(\theta)-V_{\mu_2}(\theta)\big|=\frac{|\mu_2-\mu_1|\,\sinh\theta}{(\cosh\theta+\mu_1)(\cosh\theta+\mu_2)}\ \le\ \frac{|\mu_1-\mu_2|}{2}\qquad\text{for all }\theta .
\label{eq:species}
\end{equation}
The relativity principle is therefore exact for each species and holds between species up to $O(\Delta m/M)$. For an electron and a proton and $M=10^{16}$~GeV, the residual is below $5\times10^{-17}$ for any boost. It is the kinematic counterpart of the composition-dependent free fall of Sec.~\ref{sec:curved}, $\eta\simeq\Delta m/M$, and of the relative locality of DSR models~\cite{AFKS11,AGPRR25}. All three are controlled by the same small number.

\subsection{Where the deformation becomes visible}

The fractional deviation of time dilation from special relativity is $E/M$ [Eq.~\eqref{eq:dilation}], and the bound $E<M$ is exact. Table~\ref{tab:visible} shows $E/M$ for the most energetic particles accessible today.

\begin{table}[h]
\centering
\caption{Size of the deformation, $E/M$, for representative energies ($M_{\rm Pl}=1.22\times10^{19}$~GeV).}\label{tab:visible}
\small
\begin{tabular}{@{}lccc@{}}
\toprule
Particle & $E$ & $M=10^{16}$~GeV & $M=M_{\rm Pl}$\\
\midrule
LHC proton & $6.8$~TeV & $6.8\times10^{-13}$ & $5.6\times10^{-16}$\\
Highest-energy cosmic ray~\cite{TA23} & $2.4\times10^{20}$~eV & $2.4\times10^{-5}$ & $2.0\times10^{-8}$\\
\bottomrule
\end{tabular}
\end{table}

The deformation lies beyond direct reach, as expected of a fundamental mass at high energies, but the highest-energy cosmic rays already come within five orders of magnitude of $M=10^{16}$~GeV. The bound $E<M$ also turns every observed energy into a lower limit on $M$ (Sec.~\ref{sec:constraints}).

\subsection{Summary of the demonstration}

Table~\ref{tab:checklist} lists the requirements that a relativity theory with two invariant scales must meet, and where each is established.

\begin{table}[h]
\centering
\caption{Requirements of a relativity theory with invariant $c$ and $M$.}\label{tab:checklist}
\small
\begin{tabular}{@{}>{\raggedright\arraybackslash}p{3.1cm}>{\raggedright\arraybackslash}p{7.3cm}>{\raggedright\arraybackslash}p{2.4cm}@{}}
\toprule
Requirement & How it is met & Where\\
\midrule
Relativity principle & $G_\mu\cong SO^+(1,3)$ preserves $F_\mu$; it is the Legendre image of $\Lambda_M$ & Prop.~\ref{prop:group}\\
Invariant speed $c$ & Null cone of $F_\mu$ equals the light cone of $g$ for all $\mu$; $\hat T_\Lambda=\Lambda$ on light & Props.~\ref{prop:indicatrix}, \ref{prop:group}\\
Invariant mass $M$ & $E<M$ on every mass shell, preserved by $\Lambda_M$; $E\to M$ as $v\to c$ & Sec.~\ref{sec:projective}, Table~\ref{tab:rapidity}\\
Causality & $g^F$ Lorentzian on the timelike cone, which $\hat T_\Lambda$ preserves & Props.~\ref{prop:signature}, \ref{prop:group}\\
Correspondence principle & $\mu\to0$ gives special relativity exactly; corrections are $O(E/M)$ & Tables~\ref{tab:limits}, \ref{tab:rapidity}\\
Variational dynamics & $S=-\varepsilon_0\int F_\mu\,d\lambda$, reparametrization invariant & Secs.~\ref{sec:legendre}, \ref{sec:curved}\\
Field equations & $G_{ab}(x,y)=\kappa T_{ab}(x,y)$; Einstein's vacuum equations for parallel $u$; FLRW equations exact at $y=u$ & Sec.~\ref{sec:fieldeq}\\
Curved spacetime & Navigation data become fields: $h=(1+\mu)^2g(x)$, $W=\mu u(x)/(1+\mu)$ & Prop.~\ref{prop:zermelo}, Sec.~\ref{sec:curved}\\
Agreement with experiment & All effects scale as $E/M$ or $\Delta m/M$ & Sec.~\ref{sec:constraints}\\
\midrule
Open & Dynamics of $u$; species dependence $O(\Delta m/M)$, including the choice of $\mu$ in the field equation; direction-dependent sources; action-based field equations and energy--momentum conservation & Secs.~\ref{sec:species}, \ref{sec:fieldeq}, \ref{sec:conservation}, \ref{sec:discussion}\\
\bottomrule
\end{tabular}
\end{table}

\section{Kinematics of a free particle}\label{sec:kinematics}

\begin{figure}[t]
\centering
\includegraphics[width=0.78\textwidth]{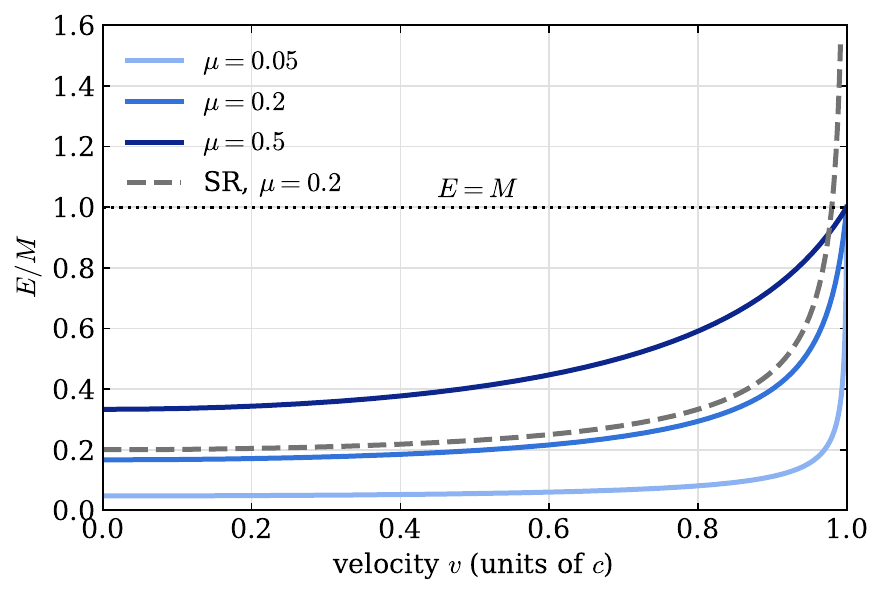}
\caption{Energy of a massive particle versus velocity, Eq.~\eqref{eq:Ep}, for three values of $\mu=m/M$. All curves saturate at $E=M$ as $v\to c$. The dashed curve is special relativity, $E=m/\sqrt{1-v^2}$, for $\mu=0.2$; it crosses $E=M$ at $v\approx0.98$.}
\label{fig:energy}
\end{figure}

We now express the kinematics through the measured velocity; Table~\ref{tab:rapidity} gives the same results in terms of rapidity. In the rest frame of $u$, with $v=|d\mathbf x/dt|$, the canonical energy and momentum $p_a=\varepsilon_0\,\partial F_\mu/\partial\dot x^a$ are
\begin{equation}
E=\varepsilon_0\,\frac{\tilde\gamma-\mu}{1-\mu},\qquad |\mathbf p|=m\,v\,\tilde\gamma,\qquad
\tilde\gamma=\frac{1}{\sqrt{1-(1-\mu^2)v^2}}.
\label{eq:Ep}
\end{equation}
As $v\to1$, $\tilde\gamma\to1/\mu$ and both $E$ and $|\mathbf p|$ tend to $M$: the speed of light is reached only in the limit where the particle's energy equals the fundamental mass (Fig.~\ref{fig:energy}).

\paragraph{Inertia versus rest energy} For $v\ll1$, $E\simeq\varepsilon_0+\tfrac12mv^2$ and $\mathbf p\simeq m\mathbf v$. The inertial mass is the Casimir mass $m$, while the rest energy is $\varepsilon_0=m/(1+m/M)$. The relation $E_{\rm rest}=mc^2$ holds only to order $\mu^0$.

\paragraph{Velocity--energy relation} Solving the canonical relations for $v$ gives
\begin{equation}
v=\frac{|\mathbf p|}{E+\dfrac{m^2}{M}\left(1-\dfrac{E}{M}\right)}.
\end{equation}
The correction to the special-relativistic $v=|\mathbf p|/E$ is of order $m^2/(ME)$ and vanishes for massless particles.

\paragraph{Time dilation} The proper time of a massive particle is the Finsler length, $d\tau=F_\mu(\dot x)\,d\lambda$. For uniform motion,
\begin{equation}
\frac{d\tau}{dt}=\frac{\sqrt{1-(1-\mu^2)v^2}-\mu}{1-\mu}=\frac{1+\mu-\mu\Gamma}{\mu+(1-\mu)\Gamma},\qquad \Gamma\equiv\frac{E}{\varepsilon_0},
\end{equation}
where $\Gamma$ is the measured ``Lorentz factor'' (energy over rest energy). To first order in $\mu$,
\begin{equation}
\frac{t_{\rm lab}}{\tau}=\Gamma+\mu(\Gamma-1)^2+O(\mu^2)\;\simeq\;\Gamma\left(1+\frac EM\right)\qquad(\Gamma\gg1).
\label{eq:dilation}
\end{equation}
At a given measured Lorentz factor $\Gamma=E/\varepsilon_0$, a particle lives longer in the laboratory than special relativity predicts, by the fraction $E/M$. If instead the comparison is made at fixed $E$ and Casimir mass $m$, i.e.\ against the special-relativistic value $E/m=\Gamma/(1+\mu)$, the fractional excess is $(E+m)/M$; the two differ only by the energy-independent term $m/M$, which is not separately observable once the rest energy $\varepsilon_0$ is measured. The proper time stops ($d\tau/dt\to0$) exactly when $E\to M$. For $M$ near the Planck scale the correction is far below laboratory sensitivity, in line with the all-order $\kappa$-Poincar\'e analysis of~\cite{AGPRR25}.

\section{Curved spacetime}\label{sec:curved}

\paragraph{Postulate} We generalize to a spacetime $(\mathcal M,g)$ carrying a unit timelike covector field $u$, $g^{ab}u_au_b=1$, by replacing $\eta$ by $g$ in Eq.~\eqref{eq:F}. The resulting $F_\mu(x,\dot x)$ is a Finsler spacetime structure for each $\mu$, and a free massive particle moves along the extremals of $S=-\varepsilon_0\int F_\mu(x,\dot x)\,d\lambda$. Photons ($\mu=0$) follow the null geodesics of $g$. In flat space $u$ is the time axis of the inertial frame in which $F_\mu$ is written, and the deformed transformations of Proposition~\ref{prop:group} leave $F_\mu$ invariant, so all inertial observers use the same expression; in curved spacetime it must become a field, either dynamical, as in Einstein--aether theory~\cite{JM01}, or fixed by the matter content, e.g.\ the cosmological rest frame.

\paragraph{Equations of motion} The Randers form~\eqref{eq:randers} gives
\begin{equation}
S=-\varepsilon_0\!\int\!\sqrt{\tilde a_{ab}\dot x^a\dot x^b}\,d\lambda+e\!\int\!u_a\dot x^a\,d\lambda,\qquad
e=\frac{\varepsilon_0\,\mu}{1-\mu}=\frac{m\mu}{1-\mu^2}\simeq\frac{m^2}{M}.
\end{equation}
The particle moves in the effective metric $\tilde a$ and couples to $u$ as a charge $e$ couples to a vector potential. With $\tilde\sigma$ the $\tilde a$-proper time and $\Omega=du$,
\begin{equation}
\varepsilon_0\,\frac{\tilde D\dot x_a}{d\tilde\sigma}=-e\,\Omega_{ab}\,\dot x^b,\qquad \Omega_{ab}=\partial_au_b-\partial_bu_a.
\end{equation}
If $u$ is closed ($\Omega=0$), the extremals are the geodesics of $\tilde a$. If moreover $u$ is parallel with respect to $g$, $F_\mu$ is a Berwald space whose geodesics are those of $g$. Deviations from Einstein's free fall are therefore sourced by $\nabla u$: its antisymmetric part (acceleration and vorticity) through the force $-e\,\Omega_{ab}\dot x^b$, and its symmetric part (expansion and shear) through the difference between the connections of $\tilde a$ and $g$. In the language of Proposition~\ref{prop:zermelo}, a curved spacetime with a fundamental mass is a Zermelo navigation problem with the sea $h=(1+\mu)^2g(x)$ and the wind field $W(x)=\mu\,u(x)/(1+\mu)$, in the spirit of the navigation description of stationary spacetimes~\cite{GHWW09}.

\paragraph{Newtonian limit} Take a static weak field, $g=(1+2\Phi)\,dt^2-(1-2\Phi)\,d\mathbf x^2$, and $u$ the static observers, $u=\sqrt{1+2\Phi}\,dt$. Expanding the Lagrangian for $v\ll1$,
\begin{equation}
L\simeq-\varepsilon_0+\tfrac12mv^2-\varepsilon_0\Phi,\qquad
\ddot{\mathbf x}=-\frac{\varepsilon_0}{m}\nabla\Phi=-\frac{\nabla\Phi}{1+\mu}.
\end{equation}
The inertial mass is $m$ and the passive gravitational mass is $\varepsilon_0$, so free fall becomes composition dependent. For two bodies $A$ and $B$,
\begin{equation}
\eta_{AB}=\frac{|a_A-a_B|}{\tfrac12|a_A+a_B|}\simeq|\mu_A-\mu_B|=\frac{|m_A-m_B|}{M}.
\label{eq:eta}
\end{equation}

\paragraph{Matter sources} Treating $u_a$ as an independent field constrained by $g^{ab}u_au_b=1$, the energy-momentum tensor of a particle is
\begin{equation}
T^{ab}(y)=\frac{\varepsilon_0(1+\mu)}{\sqrt{-g}}\int d\lambda\,
\frac{\dot x^a\dot x^b}{\sqrt{(1-\mu^2)\,g(\dot x,\dot x)+\mu^2(u\cdot\dot x)^2}}\;\delta^4\big(y-x(\lambda)\big),
\end{equation}
and the particle sources the frame field through the current
\begin{equation}
J^a(y)=-\frac{\delta S}{\delta u_a(y)}=-\frac{\varepsilon_0\,\mu}{\sqrt{-g}}\int d\lambda\,
\frac{F_\mu\,\dot x^a}{\sqrt{(1-\mu^2)\,g(\dot x,\dot x)+\mu^2(u\cdot\dot x)^2}}\;\delta^4\big(y-x(\lambda)\big).
\end{equation}
For $\mu\to0$, $T^{ab}$ reduces to the usual point-particle tensor and $J^a$ vanishes. With an Einstein--aether action for $(g,u)$ these sources close the system. A purely Finslerian alternative is the vacuum equation of Rutz~\cite{Rutz} or the action-based Finsler gravity equation~\cite{PW12,HPV19} applied to $F_\mu$; we leave the comparison of these routes to future work. Recent results constrain the Finslerian route. Einstein--Kropina solutions of the Pfeifer--Wohlfarth vacuum equation are locally flat in four dimensions~\cite{HVF26}, which bears on our $\mu\to1$ limit. The reduction of that equation to Finsler Ricci-flatness~\cite{PV26} requires a power of $F$ to be regular, with a nondegenerate metric, on the light cone; this must be checked for $F_\mu$, whose fundamental tensor degenerates there.

\paragraph{Strong fields} For static observers around a Schwarzschild black hole, the local energy of an infalling particle, $E_{\rm loc}=E/\sqrt{1-r_g/r}$, reaches $M$ at a proper distance $\rho\simeq2r_gE/M$ from the horizon. Beyond this point the projective map is singular and the particle would enter the branch $E>M$ of the Casimir, on which the group velocity exceeds $c$ everywhere. (On this branch $\pi$ lies on the past sheet, $\pi_0=-m\cosh\xi$ with $\cosh\xi>1/\mu$, and $v=\sinh\xi/(\cosh\xi-\mu)>1$ because $\xi>\operatorname{arccosh}(1/\mu)>\ln(1/\mu)$.) The divergence comes from the static frame, whose acceleration diverges at the horizon. For freely falling (Painlev\'e--Gullstrand) observers $E_{\rm loc}$ stays finite at the horizon and the problem does not arise. The choice of $u$ is therefore physical in strong fields, and only a dynamical theory of $u$ can settle it.

\section{The Einstein tensor equation in Finsler geometry}\label{sec:fieldeq}

The geometry $F_\mu$ must itself be determined by matter. We adopt the tensor form of Einstein's equation in Finsler geometry, built from the Akbar-Zadeh Ricci tensor~\cite{AZ88}, which has been used for Finsler black holes and cosmology~\cite{LC14,PDR17}. We then evaluate it exactly for $F_\mu$ on a spatially flat expanding universe, along the comoving direction $y=u$.

\subsection{Definitions}

For a Finsler function $F(x,y)$ with $L=F^2$ and $g^F_{ab}=\tfrac12\partial^2L/\partial y^a\partial y^b$, the geodesic spray, its curvature and the Finsler Ricci scalar are~\cite{BCS00}
\begin{align}
G^a&=\tfrac14\,g_F^{ab}\left(y^c\,\frac{\partial^2L}{\partial x^c\partial y^b}-\frac{\partial L}{\partial x^b}\right),\label{eq:spray}\\
R^a{}_b&=2\frac{\partial G^a}{\partial x^b}-y^c\frac{\partial^2G^a}{\partial x^c\partial y^b}+2G^c\frac{\partial^2G^a}{\partial y^c\partial y^b}-\frac{\partial G^a}{\partial y^c}\frac{\partial G^c}{\partial y^b},
\qquad \mathrm{Ric}(x,y)=R^a{}_a .\label{eq:Rab}
\end{align}
$\mathrm{Ric}$ is positively homogeneous of degree two in $y$. The Akbar-Zadeh Ricci tensor, the scalar curvature and the Finsler--Einstein tensor are
\begin{equation}
\mathrm{Ric}_{ab}=\frac12\frac{\partial^2\mathrm{Ric}}{\partial y^a\partial y^b},\qquad
S=g_F^{ab}\,\mathrm{Ric}_{ab},\qquad
G_{ab}=\mathrm{Ric}_{ab}-\tfrac12\,S\,g^F_{ab},
\label{eq:FEtensor}
\end{equation}
and the field equation reads
\begin{equation}
\boxed{\;G_{ab}(x,y)=\kappa\,T_{ab}(x,y),\qquad \kappa=8\pi G_N.\;}
\label{eq:FE}
\end{equation}
For a Riemannian metric, $\mathrm{Ric}=R_{ab}y^ay^b$, $\mathrm{Ric}_{ab}=R_{ab}$, and Eq.~\eqref{eq:FE} is Einstein's equation. By homogeneity $\mathrm{Ric}=\mathrm{Ric}_{ab}y^ay^b$, so the vacuum equations $\mathrm{Ric}_{ab}=0$ and $\mathrm{Ric}=0$ are equivalent; the latter is the vacuum equation of Rutz~\cite{Rutz} and of Li and Chang~\cite{LC14}. Away from vacuum, Eq.~\eqref{eq:FE} depends on the direction $y$. A direction-dependent source $T_{ab}(x,y)$ requires a kinetic description of matter on the tangent bundle, which we do not construct here. For a medium of particles of one species moving with the frame field, as for the cosmic fluid, we therefore impose Eq.~\eqref{eq:FE} only on the osculating direction $y=u$, following~\cite{PDR17}, with $T_{ab}(x,u)$ the perfect-fluid tensor of the comoving observers. A further restriction is that $F_\mu$ depends on the species through $\mu$, while the field equation is written for a single geometry. In what follows $\mu$ is the value for the species that dominates the source (nucleons or dark-matter particles for the matter era). A multicomponent medium, with photons on the null cone of $g$ and massive species on different $F_\mu$, needs a rule for combining the geometries, for example a $y$-averaged action~\cite{PW12,HPV19,HPV22}, and is left open. Equation~\eqref{eq:FE} is not derived from an action; the action-based field equation of Finsler gravity~\cite{PW12,HPV19} is an alternative, and Sec.~\ref{sec:conservation} shows where the two must differ.

\subsection{Closed and parallel frame fields}

\begin{proposition}[Closed frame field]\label{prop:closed}
If $du=0$, the spray of $F_\mu=\tilde\alpha+\tilde\beta$ is $G^a=\tilde G^a+P\,y^a$ with $P=e_{00}/(2F_\mu)$, and
\begin{equation}
\mathrm{Ric}=\widetilde R_{ab}\,y^ay^b+3\left(\frac{3\,e_{00}^2}{4F_\mu^2}-\frac{e_{00|0}}{2F_\mu}\right),\qquad
e_{ab}=-\frac{\mu}{1-\mu}\,\tilde\nabla_bu_a ,
\label{eq:Ricclosed}
\end{equation}
where $\tilde G^a$, $\widetilde R_{ab}$ and the stroke refer to the Lorentzian metric $\tilde a$ of Eq.~\eqref{eq:randers}, $e_{00}=e_{ab}y^ay^b$ and $e_{00|0}=e_{ab|c}\,y^ay^by^c$.
\end{proposition}
\begin{proof}
For a Randers metric whose one-form is closed, $s_{ab}=0$, the spray is $\tilde G^a+P\,y^a$~\cite{BCS00}; here $\tilde\beta_a=-\mu u_a/(1-\mu)$ and $e_{ab}=\tilde\beta_{a|b}$. For sprays related by a projective change, $\mathrm{Ric}=\widetilde{\mathrm{Ric}}+(n-1)(P^2-P_{|0})$; with $\tilde\alpha_{|0}=0$ and $\tilde\beta_{|0}=e_{00}$ one finds $P^2-P_{|0}=3e_{00}^2/(4F_\mu^2)-e_{00|0}/(2F_\mu)$, and $n=4$.
\end{proof}

\begin{corollary}[Parallel frame field]\label{cor:berwald}
If $\nabla u=0$, then $F_\mu$ is a Berwald space with the spray of $g$, $\mathrm{Ric}_{ab}=R_{ab}(g)$ and $G_{ab}=R_{ab}-\tfrac12\,g^F_{ab}\,g_F^{cd}R_{cd}$. In particular, the vacuum equations of $F_\mu$ coincide with Einstein's vacuum equations, $R_{ab}=0$, for every $\mu$. In the presence of matter the equations differ from Einstein's at order $\mu$, because $g^F_{ab}(x,y)\neq g_{ab}$; e.g.\ at $y=u$, $g^F_{ab}=(1+\mu)g_{ab}-\mu\,u_au_b$.
\end{corollary}
\begin{proof}
$\nabla u=0$ implies $\nabla\tilde a=0$, so $\tilde a$ and $g$ share the Levi-Civita connection and $e_{ab}=0$; Eq.~\eqref{eq:Ricclosed} gives $\mathrm{Ric}=R_{ab}y^ay^b$.
\end{proof}

Proposition~\ref{prop:closed} and Corollary~\ref{cor:berwald} make the statement of Sec.~\ref{sec:curved} quantitative. The deviation of the Finsler Ricci tensor from its Riemannian counterpart is sourced only by $\tilde\nabla u$, the expansion, shear and acceleration of the frame field. With matter present, there is an additional deviation of order $\mu$ from the trace and the index operations, which use $g^F$ instead of $g$. This is consistent with the result that Finsler metrics built from a vacuum solution of general relativity and a covariantly constant one-form solve the vacuum equations of Finsler gravity~\cite{Heefer24}.

\subsection{Spatially flat universe: exact evaluation along the comoving direction}

\begin{figure}[t]
\centering
\includegraphics[width=\textwidth]{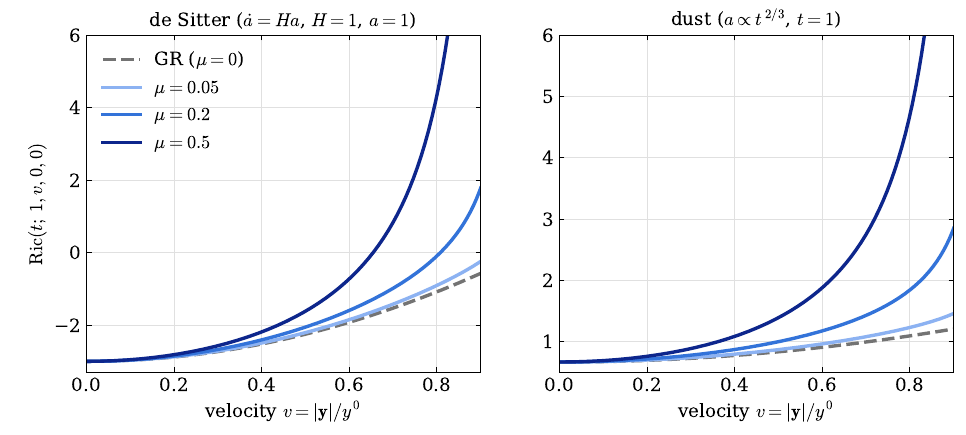}
\caption{Finsler Ricci scalar~\eqref{eq:RicFLRW} along $y=(1,v,0,0)$ for a de~Sitter universe (left) and a dust universe (right), with $a=1$ at the chosen instant, compared with general relativity (dashed). The background is the general-relativistic one, $a\propto e^{Ht}$ and $a\propto t^{2/3}$, for every $\mu$, to isolate the direction dependence; the $\mu$-dependent dust exponent of Eq.~\eqref{eq:Friedmann} changes the curves only at $O(\mu)$. All curves agree at $v=0$, i.e.\ for the comoving direction $y=u$; the Finslerian terms grow with $v$ and $\mu$ and diverge on the light cone, where $F_\mu\to0$.}
\label{fig:ricci}
\end{figure}

Take $g=dt^2-a(t)^2\,d\mathbf x^2$ and the comoving frame $u=dt$, which is closed but not parallel. Evaluating Eqs.~\eqref{eq:spray}--\eqref{eq:FEtensor} for $F_\mu$ gives, with $|\mathbf y|^2=\delta_{ij}y^iy^j$ and a dot for $d/dt$,
\begin{multline}
\mathrm{Ric}=-3\frac{\ddot a}{a}\,(y^0)^2+(1-\mu^2)\big(a\ddot a+2\dot a^2\big)|\mathbf y|^2\\
+3\left[\frac{3\mu^2(1+\mu)^2a^2\dot a^2|\mathbf y|^4}{4F_\mu^2}-\frac{\mu(1+\mu)\big(a\ddot a-3\dot a^2\big)\,y^0|\mathbf y|^2}{2F_\mu}\right].
\label{eq:RicFLRW}
\end{multline}
The first two terms are the Riemannian Ricci form $\widetilde R_{ab}y^ay^b$ of $\tilde a$; the bracket, which vanishes for $\mu=0$ and is singular on the light cone where $F_\mu\to0$, carries the genuinely Finslerian direction dependence. At $y=u$ the bracket vanishes and $\mathrm{Ric}(u)$ coincides with its general-relativistic value; away from $u$ the deviation grows with the velocity and with $\mu$ (Fig.~\ref{fig:ricci}).

\begin{figure}[t]
\centering
\includegraphics[width=\textwidth]{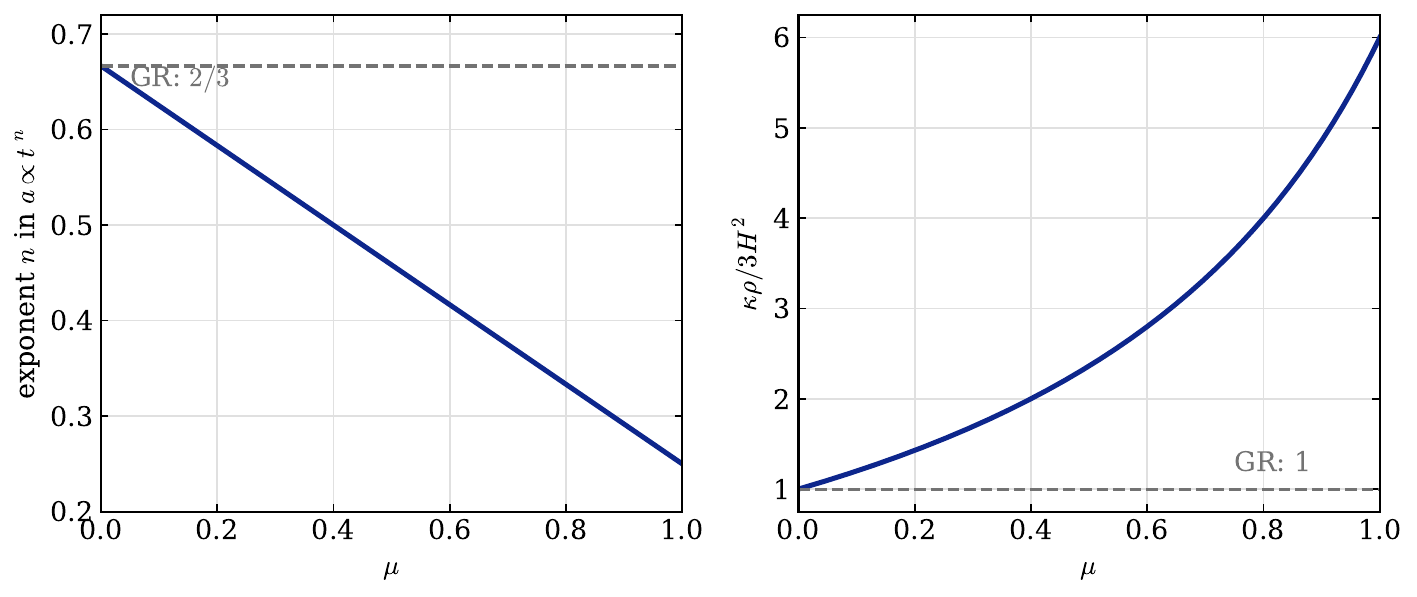}
\caption{Dust solutions of the modified Friedmann equations~\eqref{eq:Friedmann}. Left: the exponent $n$ in $a\propto t^{\,n}$; right: the ratio $\kappa\rho/3H^2$. Dashed lines give general relativity. The parameter range $0\le\mu\le1$ is shown to make the trends visible; for nucleons $\mu<2.5\times10^{-10}$.}
\label{fig:dust}
\end{figure}

At $y=u$ the Finsler metric is $g^F_{ab}(u)=\mathrm{diag}\big(1,-(1+\mu)a^2,-(1+\mu)a^2,-(1+\mu)a^2\big)$, and the Finsler--Einstein tensor~\eqref{eq:FEtensor} is diagonal. With $H=\dot a/a$, it is exactly
\begin{equation}
G^t{}_t=3H^2-\frac{15}{4}\,\mu\,\dot H,\qquad
G^i{}_j=\left[3H^2+\Big(2-\frac54\mu\Big)\dot H\right]\delta^i_j,\qquad G^t{}_i=0 .
\label{eq:GFLRW}
\end{equation}
For a perfect fluid, $T^a{}_b=\mathrm{diag}(\rho,-p,-p,-p)$, Eq.~\eqref{eq:FE} gives the modified Friedmann equations
\begin{equation}
3H^2-\frac{15}{4}\,\mu\,\dot H=\kappa\rho,\qquad
3H^2+\Big(2-\frac54\mu\Big)\dot H=-\kappa p .
\label{eq:Friedmann}
\end{equation}
Three consequences follow.
\begin{enumerate}
\item De~Sitter space, $\dot H=0$, is an exact solution for every $\mu$, with $p=-\rho$ and $3H^2=\kappa\rho$.
\item For dust, $p=0$, the second equation gives $a\propto t^{\,2/3-5\mu/12}$, and the first gives $\kappa\rho=3H^2\big(1+\tfrac{15}{8}\mu\big)+O(\mu^2)$: the effective gravitational coupling of the expansion is reduced by the fraction $15\mu/8$. The exact dependence on $\mu$, $n(\mu)=(2-\tfrac54\mu)/3$ and $\kappa\rho/3H^2=1+15\mu/(8-5\mu)$, is shown in Fig.~\ref{fig:dust}.
\item For $\mu\to0$ Eqs.~\eqref{eq:GFLRW}--\eqref{eq:Friedmann} reduce to the Friedmann equations of general relativity.
\end{enumerate}
For nucleons and $M>4\times10^{9}$~GeV (Sec.~\ref{sec:constraints}), $\mu<2.5\times10^{-10}$, so these corrections are far below cosmological sensitivity; their form, proportional to $\dot H$, is nevertheless fixed.

\subsection{Conservation law}\label{sec:conservation}

\begin{figure}[t]
\centering
\includegraphics[width=0.78\textwidth]{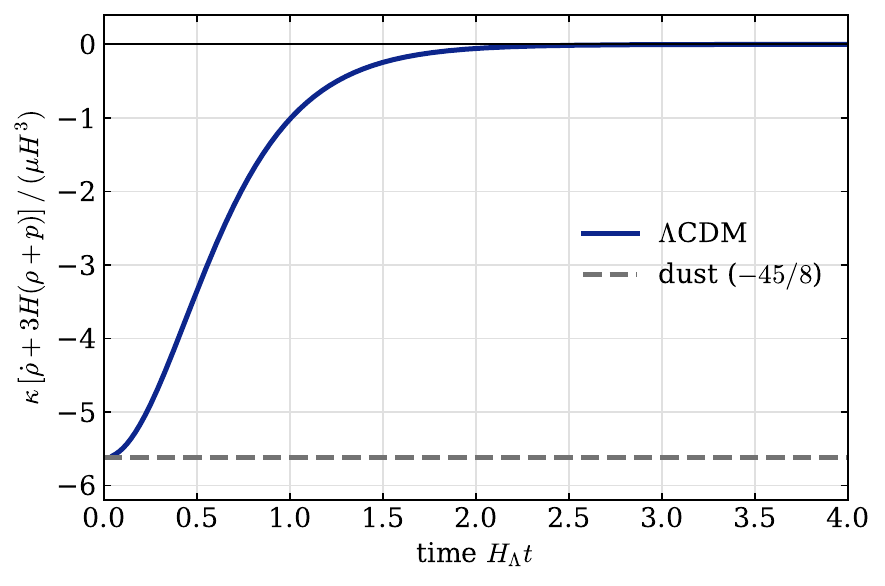}
\caption{Violation of the standard conservation law, Eq.~\eqref{eq:nonconservation}, normalized by $\mu H^3$, for a $\Lambda$CDM background with $H=H_\Lambda\coth(3H_\Lambda t/2)$. The residual tends to the dust value $-45/8$ (dashed) at early times and vanishes as the universe approaches de~Sitter space.}
\label{fig:conservation}
\end{figure}

Eqs.~\eqref{eq:Friedmann} imply
\begin{equation}
\dot\rho+3H(\rho+p)=-\frac{15\,\mu}{4\kappa}\,\frac{1}{a^2}\frac{d}{dt}\big(a^2\dot H\big).
\label{eq:nonconservation}
\end{equation}
The right-hand side vanishes for de~Sitter space and for $\mu=0$, but not in general. The tensor~\eqref{eq:FEtensor} evaluated on $y=u$ is therefore not divergence-free at order $\mu$. Either the matter exchanges energy with the frame field $u$, which then needs its own dynamics (Sec.~\ref{sec:discussion}), or the field equation must be replaced by one derived from an action~\cite{PW12,HPV19}. In that framework general covariance yields an energy--momentum conservation law averaged over the directions $y$, which reduces to the usual pointwise law for pseudo-Riemannian spacetimes~\cite{HPV22}. Equation~\eqref{eq:nonconservation} shows that for $F_\mu$ the pointwise law fails when the tensor equation is evaluated on $y=u$, so the averaged law is the natural candidate. Equation~\eqref{eq:nonconservation} gives the size of the required modification. For a $\Lambda$CDM background, $H=H_\Lambda\coth(3H_\Lambda t/2)$, the normalized residual $\kappa[\dot\rho+3H(\rho+p)]/(\mu H^3)$ interpolates between the dust value $-45/8$ at early times and zero in the de~Sitter future (Fig.~\ref{fig:conservation}).

\subsection{Computation and checks}

The results~\eqref{eq:RicFLRW}--\eqref{eq:nonconservation} were obtained analytically from Proposition~\ref{prop:closed} and verified in two independent ways.

\paragraph{Wolfram Mathematica~13} A symbolic computation in Wolfram Mathematica~13 (run time 8~s on a desktop computer) gives the following.
\begin{itemize}
\item The spray computed from its definition~\eqref{eq:spray} agrees to 30 significant digits with the closed form $\tilde G^a+e_{00}\,y^a/(2F_\mu)$ at three exact sample points, for three scale factors and $\mu=37/100$, $1/5$ and $3/5$.
\item The one-form $\tilde\beta$ is closed, and Eq.~\eqref{eq:Ricclosed} reduces identically to Eq.~\eqref{eq:RicFLRW}; for $\mu=0$ both reduce to $R_{ab}y^ay^b$ of general relativity.
\item The Akbar-Zadeh tensor and the scalar curvature at $y=u$ are
\begin{gather}
\mathrm{Ric}_{tt}(u)=-3\big(H^2+\dot H\big),\nonumber\\
\mathrm{Ric}_{ij}(u)=\tfrac12(1+\mu)\,a^2\big[6H^2+(2-5\mu)\dot H\big]\delta_{ij},\nonumber\\
S(u)=\tfrac32\big[-8H^2+(5\mu-4)\dot H\big],
\label{eq:RicU}
\end{gather}
and the Finsler--Einstein tensor equals Eq.~\eqref{eq:GFLRW} identically.
\item The density, the pressure and the conservation law~\eqref{eq:nonconservation} follow symbolically from Eqs.~\eqref{eq:Friedmann}.
\end{itemize}

\paragraph{Independent numerical check} A finite-difference evaluation of the curvature~\eqref{eq:Rab} directly from the spray, for $a(t)=1+t^2/3+\tfrac15\sin t$ and random $\mu$, $t$ and $y$, reproduces Eq.~\eqref{eq:Ricclosed} to $10^{-8}$ and Eq.~\eqref{eq:GFLRW} to $10^{-5}$, the accuracy of the finite differences.

Figures~\ref{fig:ricci}--\ref{fig:conservation} are plotted from Eqs.~\eqref{eq:RicFLRW}, \eqref{eq:Friedmann} and~\eqref{eq:nonconservation}; the same plots were also produced in Mathematica.

\section{Observational constraints}\label{sec:constraints}

The bound $E<M$ is exact, so every observed particle energy is a lower limit on $M$. The highest-energy cosmic ray~\cite{TA23} gives $M>2.4\times10^{11}$~GeV if the primary is a proton, and $M>4\times10^{9}$~GeV per nucleon if it is an iron nucleus carrying the scale $56M$ (Sec.~\ref{sec:discussion}). Stronger bounds come from the electron EDM, once the P- and CP-odd gauge sector of~\cite{CDIKM85a,Ibadov04} is included, and from free fall if atomic nuclei are the elementary carriers of $\mu$ (Table~\ref{tab:constraints}). The estimates are order-of-magnitude and assume the sensitivities quoted in the cited experiments. Deformed time dilation has recently been confronted directly with the CERN Muon Storage Ring data~\cite{LPM25}, and future muon colliders have been proposed as Planck-scale probes of it~\cite{LP21,LP24}; the $E/M$ law of Eq.~\eqref{eq:dilation} can be tested in the same way.

\begin{table}[h]
\centering
\caption{Constraints on the fundamental mass $M$.}\label{tab:constraints}
\small
\begin{tabular}{@{}>{\raggedright\arraybackslash}p{2.8cm}>{\raggedright\arraybackslash}p{4.0cm}>{\raggedright\arraybackslash}p{3.2cm}>{\raggedright\arraybackslash}p{2.4cm}@{}}
\toprule
Test & Observable in this model & Assumption & Bound on $M$\\
\midrule
Highest-energy cosmic ray~\cite{TA23} & $E<M$ for $E=2.4\times10^{20}$~eV & proton primary; iron primary with scale $56M$ & $M>2.4\times10^{11}$~GeV; $M>4\times10^{9}$~GeV\\
LHC protons & $E<M$ for $E=6.8$~TeV & --- & $M>6.8$~TeV\\
Muon storage ring, $\gamma\approx29.3$~\cite{Bailey77} & $t_{\rm lab}/\tau-\Gamma\simeq\Gamma E/M$, $E\approx3.1$~GeV & lifetime agrees with SR to $\sim10^{-3}$ & $M\gtrsim3$~TeV\\
GRB photon timing~\cite{Fermi09,LHAASO24} & none: photons follow null geodesics of $g$ & --- & no bound\\
Electron EDM~\cite{Roussy23} & $d\simeq e\ell/2=e/(2M)$ from the $\gamma^5$ vertices of the gauge sector~\cite{Ibadov04} & vertices of~\cite{CDIKM85a,Ibadov04} present at tree level; $d_e<4.1\times10^{-30}\,e\,$cm & $M\gtrsim2\times10^{15}$~GeV\\
MICROSCOPE, Pt--Ti~\cite{MICROSCOPE22} & $\eta\simeq\Delta m/M$, $\Delta m\approx137$~GeV & nuclei carry $\mu$; conservatively $\eta\lesssim10^{-14}$ & $M\gtrsim10^{16}$~GeV\\
MICROSCOPE, Pt--Ti~\cite{MICROSCOPE22} & $\eta\sim$ binding-energy differences $\times\,m_N/M$ & nucleons carry $\mu$ & negligible for $M\sim M_{\rm Pl}$\\
\bottomrule
\end{tabular}
\end{table}

\paragraph{Photons} Because the null cone of $F_\mu$ is that of $g$ for every $\mu$, there is no energy-dependent speed of light. Time-of-flight limits on Lorentz violation from Fermi-LAT~\cite{Fermi09} and LHAASO~\cite{LHAASO24}, which push linear-order scales beyond the Planck mass, therefore do not constrain $M$. This distinguishes the model from Finsler and rainbow models with modified photon dispersion~\cite{MS04,GLS07}.

\paragraph{Electric dipole moments} The free Finsler kinematics of Secs.~\ref{sec:legendre}--\ref{sec:kinematics} conserves P and CP. If the gauge sector of Kadyshevsky's theory is added, its $\gamma^5$-type vertices give leptons a dipole moment $d\simeq e\ell/2$~\cite{Ibadov04}. Taken at face value, the electron-EDM limit~\cite{Roussy23} then requires $\ell\lesssim8\times10^{-30}$~cm, i.e.\ $M\gtrsim2\times10^{15}$~GeV. This is the strongest bound once the interacting theory is included, although it depends on that sector, and it shows that the P- and CP-odd sector, not the kinematics, is where the fundamental mass is most exposed.

\paragraph{Massive neutrinos} The velocity correction of Sec.~\ref{sec:kinematics}, of relative order $m_\nu^2/(ME)$, is smaller than the standard term $m_\nu^2/(2E^2)$ by the factor $2E/M$. For PeV neutrinos and $M>4\times10^{9}$~GeV this factor is below $10^{-3}$.

\paragraph{Equivalence principle} The free-fall bound depends on what counts as ``one particle''. If the deformation is carried by the nuclei, a test of the MICROSCOPE type bounds $M$ close to the grand-unification scale. If it is carried by nucleons, composition dependence enters only through binding-energy differences and is far below current sensitivity. This dependence on the composition rule is the soccer-ball problem of Sec.~\ref{sec:discussion}.

\section{Discussion and open problems}\label{sec:discussion}

\paragraph{Physical content versus change of variables} If all observables were expressed through the auxiliary momenta $\pi$, the model would be special relativity in disguise, a known objection to DSR. In the Finsler formulation the physical clock is the Finsler length and the measured energy is the canonical one; with these identifications the time-dilation law, the velocity--energy relation and, in curved spacetime, the free-fall acceleration all differ from special and general relativity. The predictions of Secs.~\ref{sec:kinematics}--\ref{sec:constraints} are therefore operational, and they stand or fall with this identification of clocks and energies.

\paragraph{Composite bodies (soccer-ball problem)} Applied naively to a body of mass above $M$, the theory predicts absurd kinematics. A consistent rule is that a body of $N$ constituents carries the scale $NM$~\cite{MS02}, so that $\mu_{\rm body}$ equals the mass-weighted mean $\mu$ of its constituents. Which constituents are elementary decides the equivalence-principle bound of Sec.~\ref{sec:constraints}. Equation~\eqref{eq:species} shows that the residual species dependence of the frame transformations never exceeds $|\Delta\mu|/2$, so the problem is quantitatively small, but it is not removed. A derivation of this rule from a composition law for momenta~\cite{AFKS11} is needed.

\paragraph{Dynamics of the frame field} In flat space $u$ is an observer label; in curved spacetime it is a field, and the relativity principle with two invariants survives only if its dynamics respects $\Lambda_M$ locally. Coupling $F_\mu$ to Einstein--aether theory~\cite{JM01} is the most direct route. The strong-field behaviour of Sec.~\ref{sec:curved} shows that this choice is not cosmetic.

\paragraph{Relation to Kadyshevsky's theory} Kadyshevsky's anti-de~Sitter momentum space bounds the Casimir mass, $m\le M$, and keeps linear Lorentz invariance~\cite{Kadyshevsky78,KM85}. Its free mass shell is $p^2=m^2$, so the new physics sits in the interaction vertices and in the doubled field content of the five-dimensional formulation~\cite{CDIKM85a,CDIKM85b,IK89}. The projective realization used here bounds the rest energy and the total energy instead, at the price of nonlinear boosts, and puts the new physics into the free kinematics. The two pictures share a structural feature. Kadyshevsky's discrete variable $\varepsilon=\pm1$ separates normal fields from exotic ones that have no flat limit~\cite{CDIKM85b,DIKMC}; our Casimir likewise has a second branch, $E>M$, which disappears as $M\to\infty$ and on which the group velocity exceeds $c$ (Sec.~\ref{sec:curved}). Whether this branch is the classical shadow of the exotic fermion sector, and whether the P- and CP-violating vertices of the gauge sector~\cite{CDIKM85a,Ibadov04} can be written as a Finsler connection on $F_\mu$, are open questions whose answers would unify the two formulations.

\paragraph{Dynamical masses and the fundamental scale} On Kadyshevsky's hyperboloid~\eqref{eq:ads} the fifth momentum component is $p_5=\sqrt{M^2-p^2}$, which becomes $\sqrt{M^2+k^2}$ after Wick rotation. The same square-root operator appears in pseudo-Proca QED, the planar reduction of the Stueckelberg--Proca theory, where integrating out the extra coordinate gives the gauge kernel $1/\big(2\sqrt{k^2+m^2}\big)$~\cite{BAN26}. There the scale $m$ is a branch point rather than a pole, much as $M$ in Kadyshevsky's theory is a curvature radius rather than a particle mass. A Schwinger--Dyson analysis of that model shows that the scale suppresses dynamical mass generation. In the rainbow-quenched truncation the critical coupling grows as
\begin{equation}
\alpha_c(m,\Lambda)=\frac{\pi}{8}\left(1+\frac{m^2}{\Lambda^2}\right)^{3/2},
\end{equation}
and in the unquenched truncation the critical flavour number falls by about 10\% at $m/\Lambda\approx0.3$; both trends survive a Ball--Chiu vertex~\cite{BAN26}. Applying the same nonperturbative method to the gauge sector of~\cite{CDIKM85a,CDIKM85b} would test whether dynamically generated masses respect the bound $m\le M$, and whether the fundamental mass favours or suppresses chiral symmetry breaking near that bound. The analogy is formal, since the planar model is not a theory of a fundamental mass, but it shows that square-root kernels of this type can be handled with current nonperturbative tools.

\paragraph{Further work} Natural next steps are spherically symmetric solutions with a dynamical $u$, following~\cite{LPH12} and the Schwarzschild--Finsler--Randers analyses~\cite{TBKS20,KKST22}, in which the one-form is a free field while here it is fixed by the fundamental mass; kinetic theory of gases on $F_\mu$, along the lines of the Finsler--Friedmann analysis~\cite{PVFP25}; Randers--Kropina cosmology with the mass-dependent $F_\mu$~\cite{Bouali25}; quantization of the free field with the Casimir~\eqref{eq:casimir}; and a Schwinger--Dyson study of dynamical mass generation in the presence of the fundamental mass, along the lines of~\cite{BAN26}.

\section{Conclusions}

We have obtained, in closed form, the Finsler geometry dual to a relativistic mass shell with an invariant fundamental mass $M$. For a particle of mass $m$ it is the Lorentzian Randers metric $F_\mu$, $\mu=m/M$, built from the spacetime metric and a unit timelike covector. The geometry has Lorentzian signature inside the timelike cone, keeps the light cone of $g$, and interpolates between Einstein ($\mu\to0$) and Kropina ($\mu\to1$) geometry. The existence of the generalization rests on a single identity, $F_\mu(w+\mu u)=1+\mu$: the indicatrix is the Minkowski hyperboloid displaced by the wind $W=\mu u/(1+\mu)$. The relativity group is therefore the Lorentz group rotating about $W$; it is exact for each species, transforms light exactly as in Einstein's theory, and keeps both $c$ and $M$ invariant.

Its kinematics bounds every massive particle's energy by $M$ and every rest energy by $\varepsilon_0=mM/(M+m)$, and it deforms time dilation by the fraction $E/M$. In curved spacetime it predicts a universal aether charge $e\simeq m^2/M$ and a composition-dependent free fall $\eta\simeq\Delta m/M$, while leaving light propagation exactly Einsteinian. The Finsler--Einstein tensor equation has the same vacuum solutions as Einstein's equation for a parallel frame field. Evaluated along the comoving direction of an expanding universe, it yields Friedmann equations with corrections $-\tfrac{15}{4}\mu\dot H$ and $-\tfrac54\mu\dot H$ that leave de~Sitter space exact but break the standard conservation law at order $\mu$. Depending on the composition rule for bound states and on whether the P- and CP-odd vertices of Kadyshevsky's gauge sector are included, current data place $M$ above $4\times10^{9}$--$2.4\times10^{11}$~GeV from kinematics alone (iron or proton primary of the highest-energy cosmic ray), and above about $10^{15}$--$10^{16}$~GeV when the gauge sector or nuclear carriers of $\mu$ are assumed. The theory thus turns the fundamental-mass hypothesis of Kadyshevsky and Markov into a geometric generalization of relativity with definite, falsifiable consequences.

\section*{CRediT authorship contribution statement}
\textbf{Rustam Ibadov:} Conceptualization, Methodology, Formal analysis, Software, Validation, Writing -- original draft, Writing -- review \& editing, Funding acquisition.

\section*{Declaration of competing interest}
The author declares that he has no known competing financial interests or personal relationships that could have appeared to influence the work reported in this paper.

\section*{Data availability}
No data were generated in this study. The Wolfram Mathematica code used to verify the Finsler--Einstein tensor is available from the author on reasonable request.

\section*{Declaration of generative AI and AI-assisted technologies in the writing process}
During the preparation of this work the author used Claude (Anthropic) in order to improve the readability and language of the manuscript, as well as to assist with LaTeX formatting. After using this tool, the author reviewed and edited the content as needed and takes full responsibility for the content of the published article.

\section*{Acknowledgments}
The author thanks D.~G.~Pavlov for his hospitality and fruitful discussions. Funding for this study was received from the Ministry of Higher Education, Science and Innovations of the Republic of Uzbekistan through project No.~FZ-20200929385.


\end{document}